\documentclass{iau307}
\usepackage{graphicx}
\usepackage{natbib}
\usepackage{subcaption}
\usepackage{url}
\usepackage{dtklogos}
\bibpunct{(}{)}{;}{a}{}{,}

\title[Cepheid rotation] 
{Rotation and the Cepheid Mass Discrepancy}

\author[Anderson, Ekstr\"om, Georgy, Meynet, Mowlavi, and Eyer]   
{Richard I. Anderson$^1$,
  Sylvia Ekstr\"om$^1$,
  Cyril Georgy$^2$,
  Georges Meynet$^1$,
  Nami Mowlavi$^1$,
  \and
  Laurent Eyer$^1$
}

\affiliation{$^1$Observatoire de Gen\`eve, Universit\'e de Gen\`eve, 51 Ch. des
   Maillettes, 1290 Sauverny, Switzerland  \\
$^2$Astrophysics group, Lennard-Jones Laboratories, EPSAM, Keele University,
   Staffordshire, ST5 5BG, UK \\ [\affilskip] 
   email: {\tt richard.anderson@unige.ch}}

\pubyear{2014}
\volume{307} 
\pagerange{}
\jname{New windows on massive stars: asteroseismology, interferometry, and spectropolarimetry}
\editors{G. Meynet, C. Georgy, J.H. Groh \& Ph. Stee, eds.}

\begin{document}

\maketitle

\begin{abstract}
We recently showed that rotation significantly affects most observable Cepheid
quantities, and that rotation, in combination with the evolutionary status of
the star, can resolve the long-standing Cepheid mass discrepancy problem. We
therefore provide a brief overview of our results regarding the problem of
Cepheid masses. We also briefly mention the impact of rotation on the Cepheid
period-luminosity(-color) relation, which is crucial for determining
extragalactic distances, and thus for calibrating the Hubble constant.


\end{abstract}

\firstsection 
\section{Introduction}
Classical Cepheids are evolved intermediate-mass stars observed during brief
phases of stellar evolution that render them highly precise standard candles.
They are furthermore excellent laboratories of stellar structure and evolution
thanks to their variability and location in the Hertzsprung-Russell diagram.
Despite the adjectives {\it classical} and {\it standard}, Cepheids are all but
sufficiently well understood. A key symptom of this is the $45$-year-old Cepheid
mass discrepancy problem 
\citep{1968QJRAS...9...13C,1969MNRAS.144..461S,1969MNRAS.144..485S,1969MNRAS.144..511S}
that has been estimated until recently to be in the range of $10 - 20 \%$
\citep{2006MmSAI..77..207B} and has motivated much research into convective core
overshooting \citep[e.g.][]{2012ApJ...749..108P} and enhanced mass-loss
\citep{2008ApJ...684..569N}.

\begin{figure}
\centering
\begin{minipage}{0.47\textwidth}
  \includegraphics[width=\textwidth]{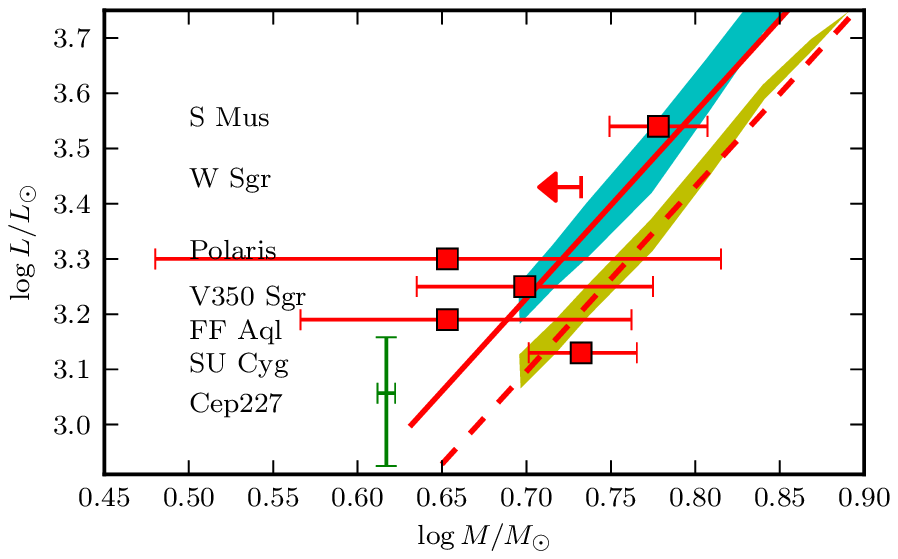}
  \caption{Mass luminosity relations of rotating models (higher $L$) better
  reproduce observed Cepheid masses than non-rotating ones, see
  \citet{2014A&A...564A.100A} for more details and references.}
  \label{fig:MLR}
\end{minipage}%
\hspace{0.5cm}
\begin{minipage}{0.47\textwidth}
  \flushright
  \includegraphics[width=\textwidth]{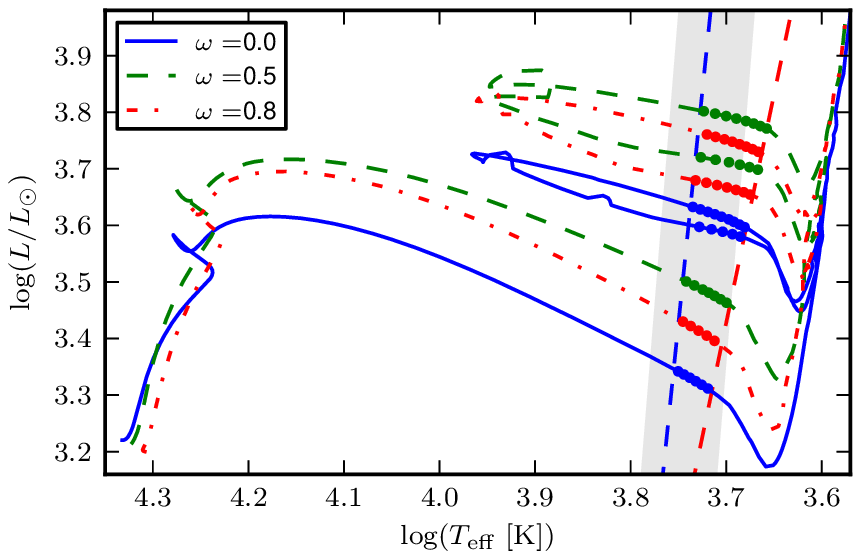}
  \caption{Evolutionary tracks for a $7\,M_\odot$ model with Solar metallicity
  and different initial rotation rates quantified by $\omega = \Omega /
  \Omega_{\rm{crit}}$. The effect of rotation during the main sequence
  carries through to the later evolutionary stages.}
  \label{fig:Models}
\end{minipage}
\end{figure}

\section{Rotation to the Rescue\label{SecOne}}

While convective core overshooting is successful in increasing core size and
thereby increasing luminosity at fixed mass, it cannot fully explain the
mass discrepancy, since high values ($\geq 20\%$ of pressure scale height) of
convective core overshooting also suppress the appearance of blue loops at the 
low-mass end, cf. \citet[Fig.~1]{2014A&A...564A.100A}. This is a problem,
since the majority of Cepheids are understood to reside on blue loops 
and have short periods, i.e., they originate from relatively low ($\sim
5\,M_\odot$) mass B-stars. 

We recently presented the first detailed investigation of the effect of rotation
on populations of classical Cepheids \citep{2014A&A...564A.100A} based on the
latest Geneva stellar evolution models
\citep{2012A&A...537A.146E,2013A&A...553A..24G} that incorporate a homogeneous
treatment of rotation over a large range of masses.
We found that rotation, together with evolutionary status (i.e., 
identification of the instability strip (IS) crossing) can resolve the mass
discrepancy, and mass-luminosity relations (MLRs) of models for typical initial
rotation rates agree better with observed Cepheid masses than models without
rotation, see Fig.\,\ref{fig:MLR}. 
Furthermore, rotation does not suppress the appearance of blue loops (cf.
Fig.~\ref{fig:Models}) and is thus in better agreement with observations than
models invoking high overshooting values.

\section{Implications\label{SecTwo}}

An important consequence of rotation is that no unique MLR applies to all stars.
The farther a star evolves along the main sequence, the larger this difference
tends to become.
The difference in main sequence turn off luminosity between
models of different rotation rates carries over into the more advanced evolutionary stages. For
Cepheids, luminosity also tends to increase between the $2^{\rm{nd}}$ and $3^{\rm{rd}}$ IS
crossings, adding further complexity. To estimate a Cepheid's mass given the
luminosity, its evolutionary status must therefore be taken into account.
Measured rates of period change provide empirical measurements of the IS
crossings, and are furthermore sensitive to initial rotation.

Finally, we point out that rotation can lead to intrinsic scatter in the
period-luminosity relation (PLR) and the period-luminosity-color-relation
(PLCR). The PLCR follows from inserting an MLR into the pulsation equation
\cite[$P \propto 1 / \sqrt{\bar{\rho}}$,][]{1879Ritter}. As there is no unique
MLR (cf. above), there cannot be a unique PLCR. This finding has potentially
important implications for the accuracy of Cepheid distances and thus for the
distance scale. Further investigation in this direction is in progress.

\bibliographystyle{iau307}
\bibliography{MyBiblio}

\end{document}